\documentclass[twocolumn,prl]{revtex4}
\usepackage{epsfig,amssymb,amsmath}
\begin{document}
\title{Variation of Longitudinal Plasma Wavelength under Irradiation and Double Resonance in Coupled Josephson Junctions}
\author{Yu. M. Shukrinov~$^{1,2}$}
\author{I. R. Rahmonov~$^{1,3}$}
\author{M. A. Gaafar~$^{1,4}$}

\address{$^{1}$ BLTP, JINR, Dubna, Moscow Region, 141980, Russia \\
$^{2}$ Max-Planck-Institute for the Physics of Complex Systems, 01187 Dresden, Germany\\
$^{3}$Umarov Physical Technical Institute, TAS, Dushanbe, 734063 Tajikistan\\
$^{4}$Department of Physics, Faculty of Science, Menoufiya University, Egypt.
}

\date{\today}

\begin{abstract}
The effect of electromagnetic wave irradiation on the phase dynamics of intrinsic Josephson junctions in high temperature superconductors is investigated. We predict three novel effects by variation  of the radiation amplitude and frequency: changing of the longitudinal plasma wavelength at parametric resonance; double resonance of the Josephson oscillations with radiation and longitudinal plasma wave; charging of superconducting layers in the current interval corresponding to the Shapiro step. The "bump" structure in IVC  recently observed experimentally is demonstrated. We also observe ragged  Shapiro steps  at double resonance.

\end{abstract}
\maketitle

Electrical and magnetic properties of intrinsic Josephson junctions (IJJ) in high temperature superconductors (HTSC) are strongly nonlinear and determined by their phase dynamics. The phase dynamics of IJJ is used to explain the mechanism of  coherent terahertz radiation which is investigated intensively  today.\cite{kadowaki12,benseman11,krasnov11,wang10,tachiki09,ozyuzer07} One of the most spectacular indications of the Josephson effect  in HTSC is locking of the Josephson oscillations of each junction to the frequency of external electromagnetic radiation. This  locking leads to appearance of steps in the current voltage characteristics (IVC) at quantized voltages $V_n=n\hbar\omega/2e$, called Shapiro steps (SS).  Here $\omega$ is the frequency of the applied signal, $\hbar$ - the Planck constant, $e$ -  the elementary charge; $n$ - integer number.\cite{Shapiro, Tinkham}

Another interesting feature of IJJ is a longitudinal plasma wave (LPW) propagating along the c axis.\cite{koyama96,kleiner2000} It follows from the fact that the thickness of the S-layers is comparable to the Debye screening length $r_D$, and so there is no complete screening of electric charge in a separate S-layer. The frequency of Josephson oscillations $\omega_J$ is determined by the voltage in the junction, and  at $\omega_J=2 \omega_{LPW}$ ($\omega_{LPW}$ is LPW frequency) the parametric resonance (PR) is realized: the Josephson oscillations excite the LPW by their periodical actions.  IVC of IJJ demonstrate the multiple branch structure\cite{sm-sust07,sm-prl07,smp-prb07,sg-prb11}    and  have a breakpoint (BP) and some breakpoint region (BPR) in the outermost branch before transition to the inner branch.

As it is known, the one-dimensional models with coupling between junctions capture the main features of real IJJs, like hysteresis and branching of the IVC, and help to understand their physics. An interesting and very important fact is that the 1D models can also be used to describe the properties of a parallel array of Josephson junctions, which is often considered as a model for long Josephson junctions. In particular, in Refs.\onlinecite{pfeiffer06,pfeiffer08} the experimental data have demonstrated a series of resonances in the IVC of the array and were analyzed using the discrete sine-Gordon model and an extension of this model
which includes a capacitive interaction between neighboring Josephson junctions. The parametric instabilities of an one-dimensional parallel array of $N$
identical Josephson junctions were predicted by theoretical analysis of the discrete sine-Gordon equation (also known as the Frenkel-Kontorova model)  and observed
experimentally in Ref.\onlinecite{watanabe95}. In particular, the novel resonant steps related to the parametric instability were found in the IVC of
discrete Josephson ring. It was verified experimentally that such steps occur
even if there were no vortices in the ring.

In this letter we investigate the effect of electromagnetic irradiation on the phase dynamics of IJJs and  temporal oscillations of the electric charge in superconducting layers. The results carry with them several important messages concerning the variation of the wavelength of the LPW, the charging of the S-layers and their  effects on IVC of the coupled JJ. We investigate the influence of the external radiation on the  PR by: i) increase of amplitude of radiation at fixed frequency ; ii) decrease of frequency of radiation which brings SS to the parametric resonance region (PRR).
An irradiation leads to the decrease of the hysteresis in IVC of IJJ\cite{kleiner-book,prb08}, so it's expected that the PR point ($\omega_J=2\omega_{LPW}$) would be shifted also, and LPW frequency would increased . As far as we know such study have not been done yet.

To investigate the phase dynamics of IJJ we use one-dimensional  the CCJJ+DC model\cite{sms-physC06} where the gauge-invariant phase differences $\varphi_l(t)$  between S-layers $l$ and $l+1$ in the presence of electromagnetic irradiation is described by the system of equations:
\begin{equation}
\label{syseq} \left\{\begin{array}{ll} \displaystyle\frac{\partial \varphi_{l}}{\partial
t}=V_{l}-\alpha(V_{l+1}+V_{l-1}-2V_{l})
\vspace{0.2 cm}\\
\displaystyle \frac{\partial V_{l}}{\partial t}=I-\sin \varphi_{l}-\beta\frac{\partial \varphi_{l}}{\partial t} + A\sin\omega t + I_{noise}
\end{array}\right.
\end{equation}
where $t$ is dimensionless time, normalized to the inverse plasma frequency $\omega^{-1}_p$ , where $\omega_{p}=\sqrt{2eI_c/\hbar C}$,
$\beta=1/\sqrt{\beta_{c}}$, $\beta_{c}$-McCumber parameter, $\alpha$  gives the coupling between junctions\cite{koyama96},
 $A$ is the amplitude of the radiation.
To find the IVC of the stack of IJJ we solve this system of nonlinear second-order differential equations (1) using the fourth order Runge-Kutta method. In our simulations we measure the voltage in units of $V_0=\hbar\omega_p/(2e)$, the frequency in units of $\omega_{p}$, the bias current $I$ and the amplitude of radiation $A$ in units of $I_c$. To study the time dependence of the electric charge in the S-layers, we use the Maxwell equation
$\emph{div} (\varepsilon\varepsilon_0 E) = Q$, where $\varepsilon$ and
$\varepsilon_0$ are relative dielectric and electric constants. The charge density $Q_l$ in the S-layer $l$ is proportional to the difference between the voltages $V_{l}$ and $V_{l+1}$ in the neighbor insulating layers $Q_l=Q_0 \alpha (V_{l+1}-V_{l})$,
where $Q_0 = \varepsilon \varepsilon _0 V_0/r_D^2$.   We have taken even number of junctions in the stack $N=10$ to escape the additional modulations of the electric charge oscillations in the BPR which appear in case of odd number of junctions and see clear effect of radiation only. Numerical calculations have been done for a stack with the coupling parameter $\alpha = 0.05$, dissipation parameter $\beta= 0.2$ and periodic boundary conditions. We note that the results are  not very sensitive to the  parameter values and boundary conditions  in their wide region.  The details of the model and simulation procedure are presented in Ref.\onlinecite{smp-prb07}

It is known that in case of single JJ the increase in irradiation amplitude $A$ decreases the
hysteresis region, i.e., it leads to the decrease of the critical current value and the increase of the return
current $I_R$.\cite{kleiner-book} For the stack of coupled JJ the external radiation leads additionally to the series of novel effects related to the parametric resonance and creation of the LPW  propagating along the c axis.\cite{koyama96,kleiner2000} We will describe below the changing of LPW wavelength, additional resonances around SS and  double resonance $\omega_J=\omega=2\omega_{LPW}$ with increase in the amplitude of radiation $A$.
\begin{figure}[htb]
 \centering
\includegraphics[height=55mm]{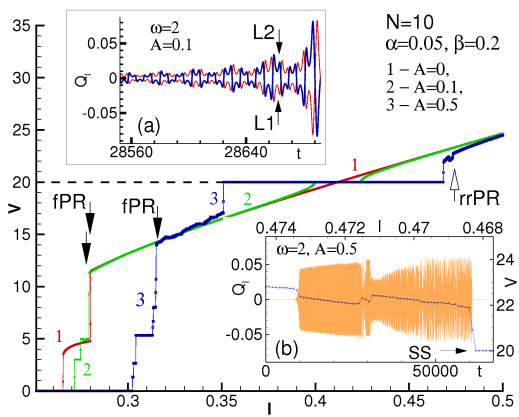}
\caption{IVC of a stack with 10 coupled JJ without irradiation (curve 1) and under radiation with frequency $\omega=2$ and amplitude $A=0.1$ (curve 2) and amplitude $A=0.5$ (curve 3). Insets (a) and (b) illustrate the modulation of the charge oscillations  at $A=0.1$ and the "bump" structure in IVC at $A=0.5$, relatively.} \label{1}
\end{figure}

First we describe the case $\omega>2\omega_{LPW}$, when the SS is above the PRR in IVC. Fig.~\ref{1} show three IVC of a stack with 10 coupled JJ: without irradiation (curve 1) and under radiation with $\omega=2$, $A=0.1$ (curve 2) and  $A=0.5$ (curve 3). At $\omega=0$ PR is characterized  by breakpoint current $I_{bp}\simeq0.28$ and breakpoint voltage $V_{bp}\simeq11.51$, corresponding to the Josephson frequency $\omega_{J}\simeq1.151$.\cite{sm-sust07} With increase in amplitude of irradiation the PRR in IVC is shifted up along the voltage axis. As we can see, the first Shapiro step is developed on the outermost branch of IVC in the  hysteresis region at $V=\omega_J * N = 20$. Dashed line stresses this fact.  Arrows indicate the position corresponding to the appearance of PR in the stack. Hollow arrow indicate the additional PR before SS which we will discuss below.

Inset  (a) shows time dependence of the electric charge in two superconducting layers (L1 and L2) at  $A=0.1$. Instead of monotonic exponential  increase of charge  observed in case without radiation, we see exponential but modulated growth.
The charges on the neighbor layers are equal in magnitude and opposite in sign. This is true for all adjacent layers and corresponds to the $\pi$-mode, so the $\pi$-mode is survived under radiation with $A=0.1$. Results of FFT analysis (not presented here) of the time dependence of voltage V(t) in each JJ and charge $Q(t)$ in each $S$-layer  show that this modulation is due to beating between the external and Josephson frequencies.

The irradiation can change the character of the charge-time dependence essentially and bring  about a "bump" structure on  the outermost branch of IVC, as shown in the inset (b) at $\omega=2$ and  $A=0.5$. At these parameters, as was mentioned above, the additional PR appears in the stack before Shapiro step (shown by hollow arrow in Fig.~\ref{1}). We call this resonance  a radiation related PR (rrPR) to distinguish it from a fundamental PR (fPR) manifested  also without radiation.   The bump structure in IVC were recently observed experimentally.\cite{ozyuzer-unpub,benseman11,iso-apl08} It is likely that the charging of the S-layers  appears not only at parametric resonance, but also at other types of resonances in coupled JJs. \cite{krasnov11,koshelev10,lin11} This likelihood has not yet been investigated in detail.
\begin{figure}[htb]
 \centering
\includegraphics[height=55mm]{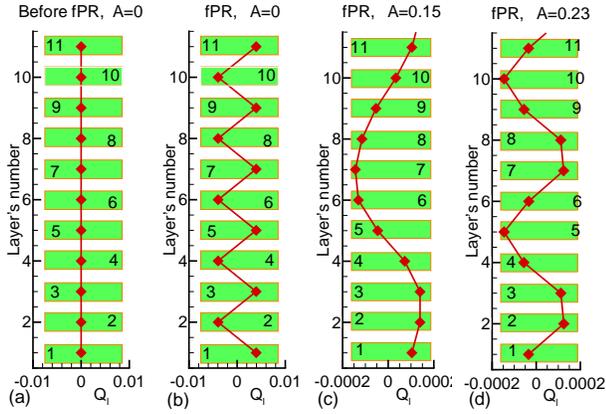}
\caption{Demonstration of changing of LPW wavelength with increase of the amplitude of radiation. The numbers count the layers in the stack.} \label{2}
\end{figure}

Fig.~\ref{2} demonstrates the effect of the amplitude increasing at $\omega=2$  on the wavelength of the fPR LPW. We show that before the  resonance region in IVC (Fig.~\ref{2}a) the charge in the layers is zero (in noise level). In the growing region of the resonance (Fig.~\ref{2}b) the amplitude of the charge oscillations increases exponentially forming the LPW with $k=\pi$ ($\lambda=2d$). At $A=0.14$ \emph{the wavelength of the created LPW at the parametric resonance is changing  by the external radiation.} The charge distribution along the stack, presented in Fig.~\ref{2}(c), illustrates the wave with $\lambda=10d$. At $A=0.23$ we fixed additional changes of the LPW: $\lambda=10d  \Rrightarrow \lambda=5d$.

\begin{figure}[htb]
 \centering
\includegraphics[height=35mm]{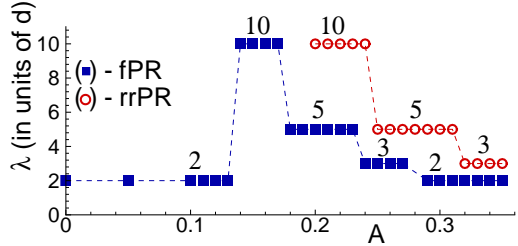}
\caption{ Demonstration of changing of LPW wavelength at fundamental PR (filled squares) and radiation related PR with increase of the amplitude of radiation at $\omega=2$.} \label{3}
\end{figure}

Results of detailed investigations of the irradiation effects at $\omega=2$ in the amplitude range $(0,0.35)$ are summarized  in Fig.~\ref{3} which shows the variation of the LPW wavelength with $A$. In the fundamental PR  we register the following transitions of LPW with increase in A: $\lambda=2d \Rrightarrow \lambda=10d \Rrightarrow \lambda=5d \Rrightarrow \lambda=3d \Rrightarrow \lambda=2d$. As we will discuss below, the increase in $A$ leads to the appearance of an additional PR before SS. The increase in $A$ in the interval $0\leq A \leq 0.35$   also changes the LPW at this radiation related resonance: $\lambda=10d \Rrightarrow \lambda=5d \Rrightarrow \lambda=3d$.

\begin{figure}[htb]
 \centering
\includegraphics[height=55mm]{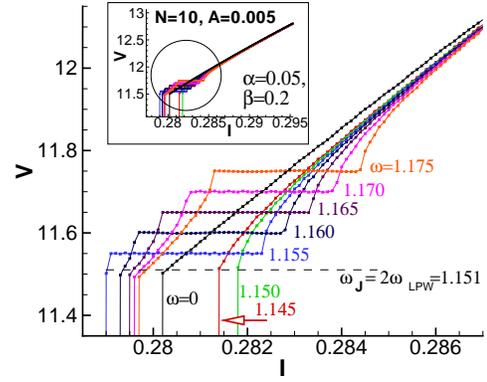}\vspace{1cm}
\caption{IVC of a stack with 10 coupled JJ  under radiation with amplitude $A=0.005$  and different frequencies. Thick curve (black online) shows IVC without irradiation. Inset stresses the coincidence of the curves before Shapiro step.
} \label{4}
\end{figure}

\begin{figure}[htb]
 \centering
\includegraphics[height=55mm]{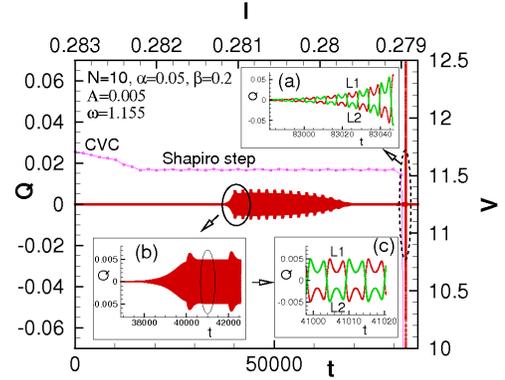}
\caption{Demonstration of the Shapiro step "charging": charge oscillations in SS current interval.  Inset (a) enlarges the charge oscillations in the parametric resonance region, insets (b) and (c) enlarges in consecutive order the charge oscillations in SS region.
} \label{5}
\end{figure}
The double resonance condition $\omega_J=\omega=2\omega_{LPW}$ can be approached by decreasing of  the radiation frequency.
In Fig.~\ref{4} we show the IVCs of a stack with 10 coupled JJ  under radiation with amplitude $A=0.005$  and different frequencies. The numbers near the corresponding curves indicate the frequency of external radiation. Thick curve (black online) shows IVC without irradiation. Inset stresses the coincidence of the all curves before Shapiro steps. The SS does not appear at frequency smaller than $\omega=1.151$, because jump to another branch is happened.

The approaching to the double resonance  demonstrates an interesting feature of coupled JJ which is absent in case of single JJ: when external frequency is close enough to the PR  condition $\omega_J=2\omega_{LPW}$, then \emph{charge oscillations appear in S-layers in current interval
corresponding to the Shapiro step ("charging of SS").} In our case such charging of SS appears starting from $\omega\simeq1.1555$, while for fPR without radiation $\omega_J=1.151$. The amplitude of oscillations and current interval of charging ("width of charging") is growing with approaching of double resonance.  Fig.~\ref{5} demonstrates the charging of SS at $\omega=1.155$. Charge oscillations in $S$-layers correspond to the $\pi$-mode of LPW. The enlarged parts of charge-time dependence are shown in consecutive order  in the insets (b) and (c).  The  inset (a) enlarges the charge oscillations in the time interval, correspond to the bias current close to the transition to inner branch and demonstrates that the fPR is survive at these radiation parameters. It also corresponds to the creation of the $\pi$-mode of LPW.
\begin{figure}[htb]
 \centering
\includegraphics[height=55mm]{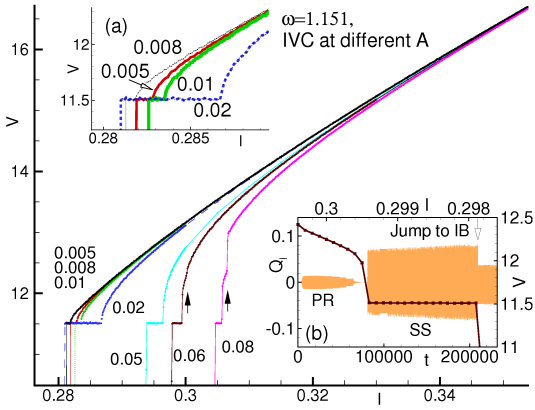}
\caption{IVC of a stack with 10 JJ at $\omega=1.151$ and different  amplitude of radiation. Inset (a) enlarges the part with small $A$. Inset (b) shows the charge-time dependence at $A=0.06$.} \label{6}
\end{figure}

\begin{figure}[htb]
 \centering
\includegraphics[height=30mm]{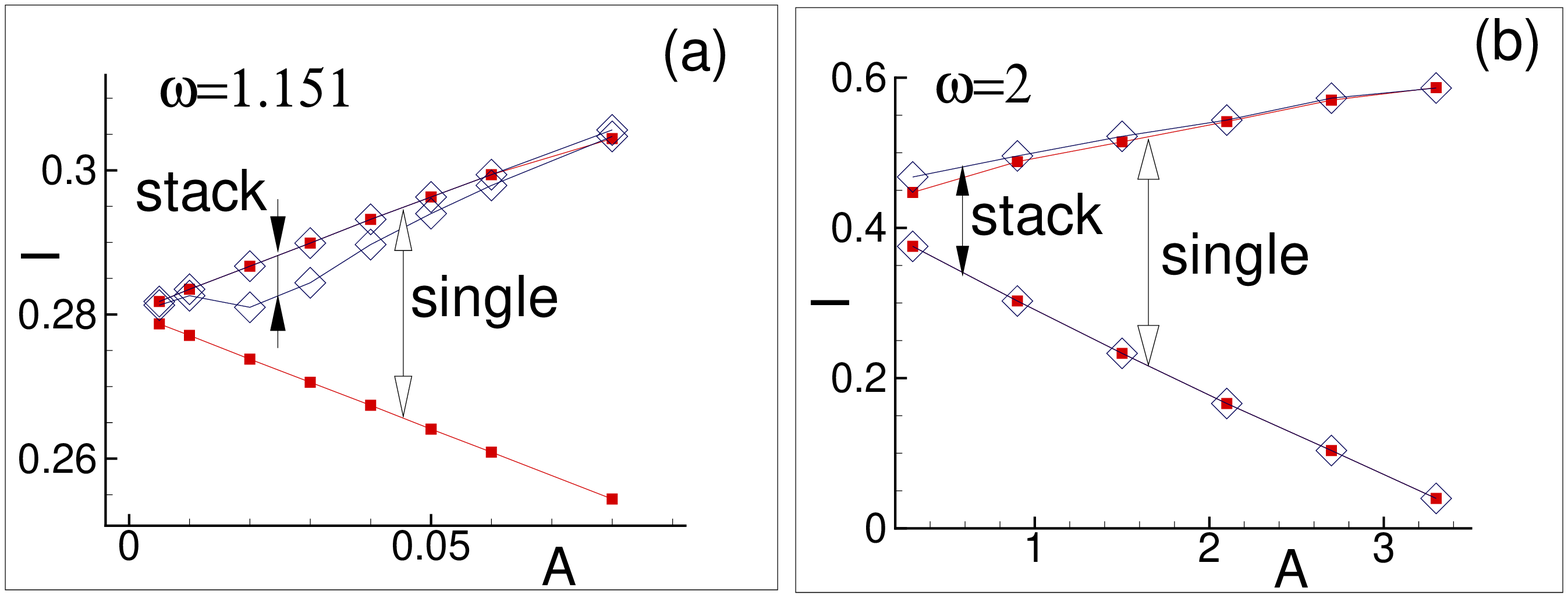}
\caption{(a) Demonstration of SS width changing for single JJ and stack of IJJ at $\omega=1.151$ (b) the same at $\omega=2$} \label{7}
\end{figure}

Fig.~\ref{6} shows the effect of amplitude increasing at $\omega=1.151$, i.e. at double resonance conditions.  Inset (a) enlarges the part of figure with small $A$. In this case even small amplitude of radiation leads to the "charging" of Shapiro steps. We see it in the inset (b), where  the charge-time dependence at $A=0.06$ together with IVC are presented. At double resonance the transition to inner branch is directly happened from the SS. In this case the Bessel function dependence of the SS width on $A$ for coupled JJ is broken and we  observe \emph{a "ragged" SS}. To stress this effect, we show in Fig.~\ref{7} A-dependence of the bias current at the beginning and at the end of SS for single JJ (hollow diamond) and stack of 10 JJ (filled squares) at $\omega=1.151$ (Fig.~\ref{7}a) and $\omega=2$ (Fig.~\ref{7}b). Double arrows indicate the corresponding SS width. We see that at $\omega=1.151$ the value of SS width is cut off ("ragged") in compare with the cases of single JJ and stack at $\omega=2$, when SS if far from fundamental PR.

An increase of the radiation amplitude leads to the appearance of the PR before SS (rrPR, $I>I_{SS}$) and  the creation of LPW with  wavelength $\lambda=5d$.  Charged and ragged SS is still conserved. This situation is demonstrated in  inset (b) of Fig.~\ref{6} which shows the charge-time dependence together with IVC at $A=0.06$. We see two charged regions: the rr-PR region and SS region. Appearance  of rr-PR in the system  before SS is reflected in IVC by its deformation (appearance of the breakpoint). We indicate such deformation  by arrows at $A=0.06$ and $A=0.08$. Further increase of radiation's amplitude leads to disappearance of the main SS. We note that we observe a "charging" of some SS harmonics as well. Detailed description of this phenomena and various manifestation of the double resonance in coupled JJ will be considered somewhere else.\cite{prb2012preparation}

As summary, we demonstrated a series of novel effects  of external electromagnetic irradiation on the phase dynamics of coupled Josephson junctions and their current voltage characteristics which are absent in case of single junction.  The results carry with them several important messages about the variation of the wavelength of the longitudinal plasma wave with changing the amplitude of external radiation, the charging of the S-layers in current interval corresponding to the Shapiro step and double resonance of the Josephson oscillations with irradiation and longitudinal plasma wave. These results should be taken into account in further research and applications concerning behavior of the intrinsic JJ in HTSC under external electromagnetic radiation.

We thank A. E. Botha, P. Seidel and M. Suzuki for fruitful discussion of some results of this paper.


\begin{thebibliography}{10}
\bibitem{kadowaki12}  M. Tsujimoto {\em et al.}, Phys. Rev. Lett. {\bf 108}, 107006 (2012).
\bibitem{benseman11} T. M. Benseman {\em et al.},  Phys. Rev. B {\bf 84}, 064523 (2011).
\bibitem{krasnov11} V. M. Krasnov, Phys. Rev. B {\bf 83}, 174517 (2011).
\bibitem{wang10} H. B. Wang {\em et al.}, Phys. Rev. Lett. {\bf 105}, 057002 (2010).
\bibitem{tachiki09} M. Tachiki, S. Fukuya, and T. Koyama, Phys. Rev. Lett. {\bf 102}, 127002 (2009).
\bibitem{ozyuzer07} L. Ozyuzer {\em et al.}, Science {\bf 318}, 1291 (2007).
\bibitem{Shapiro}S. Shapiro, Phys. Rev. Lett. {\bf 11}, 80 (1963)
\bibitem{Tinkham}M. Tinkham, Introduction to Superconductivity, 2nd edition,
        McGraw-Hill, New York (1996)
\bibitem{koyama96}T. Koyama and M. Tachiki, Phys. Rev. B {\bf 54}, 16183 (1996)
\bibitem{kleiner2000}R. Kleiner, T. Gaber, and G. Hechtfischer, Phys. Rev. B {\bf 62}, 4086 (2000)

\bibitem{sg-prb11}Yu. M. Shukrinov and M. A. Gaafar, Phys. Rev. B {\bf 84}, 094514 (2011).
\bibitem{sm-sust07} Yu. M. Shukrinov, F. Mahfouzi, Supercond. Sci. Technol., {\bf 19}, S38-S42 (2007).
\bibitem{sm-prl07} Yu. M. Shukrinov, F. Mahfouzi, Phys. Rev. Lett. {\bf 98}, 157001 (2007).
\bibitem{smp-prb07}Shukrinov, Yu., Mahfouzi, F., Pedersen, N.: Phys.Rev. B, {\bf75}, 104508 (2007)
\bibitem{pfeiffer06} J. Pfeiffer {\em et al.}, Phys.Rev.Lett, {\bf96}, 034103 (2006).
\bibitem{pfeiffer08} J. Pfeiffer {\em et al.}, Phys.Rev.B, {\bf77}, 024511 (2008).
\bibitem{watanabe95}  S. Watanabe, S. H. Strogatz {\em et al.}, Phys. Rev. Lett. {\bf 74}, 379 (1995).
\bibitem{kleiner-book} W. Buckel and R. Kleiner, {\em Superconductivity: Fundamentals and Applications} (Wiley-VCH, 2004).
\bibitem{prb08}Yu. M. Shukrinov, F. Mahfouzi, M. Suzuki, Phys. Rev. B {\bf 78}, 134521 (2008).
\bibitem{sms-physC06} Yu. M. Shukrinov, F. Mahfouzi, P. Seidel.  Physica C {\bf 449}, 62 (2006).
\bibitem{iso-apl08} A. Irie, Yu. M. Shukrinov, and G. Oya, Appl. Phys. Lett. {\bf 93}, 152510 (2008).
\bibitem{ozyuzer-unpub}F. Turkoglu {\em et al.}, unpublished.
\bibitem{lin11} S.-Z. Lin, X. Hu, and L. Bulaevskii, Phys. Rev. B {\bf 84}, 104501 (2011).
\bibitem{koshelev10} A. E. Koshelev, Phys. Rev. B {\bf 82}, 174512 (2010).
\bibitem{prb2012preparation}Yu. M. Shukrinov, I. R. Rahmonov, M. A. Gaafar, Phys. Rev. B in preparation.
\end{thebibliography}
\end{document}